# Wigner and his many friends: A new no-go result?

Sebastian Fortin and Olimpia Lombardi

*University of Buenos Aires, CONICET*

## 1.- Introduction

In April 2016, Daniela Frauchiger and Renato Renner published online an article entitled "Single-world interpretations of quantum theory cannot be self-consistent", in which they introduced a *Gedankenexperiment* that led them to conclude that, if "quantum theory is applied to model an experimenter who herself uses quantum theory", then "no single-world interpretation can be logically consistent." (Frauchiger and Renner 2016: 1). That argument intended to support the many-worlds interpretation of quantum mechanics, to the extent that it would force us "to give up the view that there is one single reality." (Frauchiger and Renner 2016: 22). In a new version of the paper, now entitled "Quantum theory cannot consistently describe the use of itself" and published in *Nature Communications* in September 2018, the authors moderate their original claim. In this new version, the same *Gedankenexperiment* is proposed to "investigate the question whether quantum theory can, in principle, have universal validity", and the conclusion is "that quantum theory cannot be extrapolated to complex systems, at least not in a straightforward manner." (Frauchiger and Renner 2018: 1); on this basis, the authors consider how the different interpretations of standard quantum mechanics and the different quantum theories should face their result.

Since its first online publication, the Frauchiger and Renner (F-R) argument has caused quite a splash in the field of quantum foundations. In general, it has been considered as a *new* no-go result for quantum mechanics. For instance, in the website of the Perimeter Institute of Theoretical Physics one can find a video of the talk entitled "Frauchiger-Renner no-go theorem for single-world interpretations of quantum theory", given by Lidia del Rio (2016) only two months after the original publication, in June 2016. But, in many cases, more extreme reactions can be found, based on conceiving the F-R argument as a kind of proof of the *inconsistence* of quantum mechanics. This idea, for instance, is suggested by a post of the Department of Physics of the ETH Zürich (the university to which Frauchiger and Renner belong), motivated by the recent publication of the paper; that post, entitled "Searching for errors in the quantum world" (Würsten 2018), asks "How is it possible for a theory to be inconsistent when it has repeatedly been so clearly confirmed by experiments?" (the post is reproduced in the website of *Science Daily*). In turn, with the title "Reimagining of Schrödinger's cat breaks quantum mechanics —and stumps physicists" (Castelvecchi 2018), an article appeared in the section "News" of *Nature* (the article is reproduced in *Scientific American*). And if one does not restrict the attention to highly reputed journals and websites, it turns to be impossible to keep track of the huge number of comments to the new result in other websites and personal blogs.

The immense impact of Frauchiger and Renner's work is due to the fact that their argument is neutral regarding interpretation: on the basis of three very generic and seemingly reasonable assumptions that do not include interpretive premises, the argument leads to a contradiction. This fact is viewed as pointing to a deep shortcoming of quantum mechanics itself, which contrasts with the extraordinary success of the theory.

In this article we will focus on the published version of the paper. Our purpose is not to consider and analyze all the comments of Frauchiger and Renner's work since it was proposed, because this would be an unattainable task. Our aim is to offer a careful reconstruction of the F-R argument, which in general is not elucidated with sufficient detail in the many debates about its assumptions and scope. Such a reconstruction will allow us to show that: (i) the argument can be more clearly formulated with no reference to what subjects know or see, but rather only in terms of quantum propositions, (ii) in contrast to what some commentators suppose, the argument does not require the hypothesis of collapse to arrive to its conclusion, and (iii) the contradiction resulting from the F-R argument is inferred by making classical conjunctions between different and incompatible contexts. On the basis of this clarification, we will finally argue that the conclusion of the F-R argument is not as novel and original as its great impact might make us to suppose.

## 2.- The experimental setup and a first approach to the argument

The *Gedankenexperiment* proposed in Frauchiger and Renner's article is a sophisticated reformulation of Wigner's friend experiment (Wigner 1961). In that original thought experiment, Wigner considers the superposition state of a particle in a closed laboratory where his friend is confined. When Wigner's friend measures the particle, the state collapses to one of its components. However, from the outside of the laboratory, Wigner still assigns a superposition state to the whole composite system particle+friend+laboratory.

The F-R argument relies on duplicating Wigner's setup. Let us consider two friends $F_1$ and $F_2$ located in separate and isolated labs $L_1$ and $L_2$ respectively, where the labs are represented by the Hilbert spaces $\mathcal{H}_1$ and $\mathcal{H}_2$ respectively.[1] $F_1$ measures the observable $C$ of a biased "quantum coin" in the state $(1/\sqrt{3})|h\rangle + (\sqrt{2/3})|t\rangle$, where $|h\rangle$ and $|t\rangle$ are the eigenstates of $C$, and $h$ and $t$ are its respective eigenvalues. $F_1$ prepares a qubit in the state $|\downarrow\rangle$ if the outcome is $h$, or in the state $|\rightarrow\rangle$ if

---
[1] We slightly modify the original terminology for clarity.

the outcome is *t*, and sends it to $F_2$. When $F_2$ receives the qubit, she measures its observable $S_z$. After these two measurements, the state of the whole system composite of the two labs is:

$$|\Psi\rangle = \frac{1}{\sqrt{3}}|H\rangle|\Downarrow\rangle + \sqrt{\frac{2}{3}}|T\rangle|\Rightarrow\rangle \in \mathcal{H}_1 \otimes \mathcal{H}_2 \qquad (1)$$

where

- $|H\rangle$ and $|T\rangle$, eigenstates of an observable *A* with eigenvalues *H* and *T*, are the states of the entire lab $L_1$ when the outcome of $F_1$'s measurement is *h* and *t*, respectively,

- $|\Uparrow\rangle$ and $|\Downarrow\rangle$, eigenstates of an observable *B* with eigenvalues $\Uparrow$ and $\Downarrow$, are the states of the entire lab $L_2$ when the outcome of $F_2$'s measurement is $+1/2$ and $-1/2$, respectively,

- and $|\Rightarrow\rangle = 1/\sqrt{2}\left(|\Uparrow\rangle + |\Downarrow\rangle\right)$.

The *Gedankenexperiment* continues by considering two "Wigner" observers, $W_1$ and $W_2$, located outside the labs, who will respectively measure the observables *X* and *Y* of labs $L_1$ and $L_2$:

- *X* has eigenvalues $\text{fail}_X$ and $\text{ok}_X$, with respective eigenvectors $|\text{fail}_X\rangle$ and $|\text{ok}_X\rangle$ such that:

$$|\text{fail}_X\rangle = \frac{1}{\sqrt{2}}(|H\rangle + |T\rangle) \qquad |\text{ok}_X\rangle = \frac{1}{\sqrt{2}}(|H\rangle - |T\rangle) \qquad (2)$$

- *Y* has eigenvalues $\text{fail}_Y$ and $\text{ok}_Y$, with respective eigenvectors $|\text{fail}_Y\rangle$ and $|\text{ok}_Y\rangle$, such that:

$$|\text{fail}_Y\rangle = \frac{1}{\sqrt{2}}(|\Downarrow\rangle + |\Uparrow\rangle) \qquad |\text{ok}_Y\rangle = \frac{1}{\sqrt{2}}(|\Downarrow\rangle - |\Uparrow\rangle) \qquad (3)$$

Before analyzing the consequences of the experiment, Frauchiger and Renner point out that their argument can be conceived as a no-go theorem that proves that three "natural-sounding" assumptions, (Q), (C), and (S), cannot all be valid (2018: 2):

(Q) *Compliance with quantum theory*: Quantum mechanics is universally valid, that is, it applies to systems of any complexity, including observers. Moreover, an agent knows that a given proposition is true whenever the Born rule assigns probability 1 to it.

(C) *Self-consistency*: Different agents' predictions are not contradictory.

(S) *Single-world*: From the viewpoint of an agent who carries out a particular measurement, this measurement has one single outcome.

In the 2016 paper, Frauchiger and Renner implicitly consider (Q) and (C) as unavoidable: as a consequence, they claim that their argument shows that "no single-world interpretation can be logically consistent" (2016: 1) and, therefore, "we are forced to give up the view that there is one single reality" (2016: 22). By contrast, in the 2018 paper, they stress that "[t]he theorem itself is neutral in the sense that it does not tell us which of these three assumptions is wrong" (2018: 2); as a consequence, they admit the possibility of different theoretical and interpretive viewpoints regarding their result, and include a table that shows which of the three assumptions each interpretation or quantum theory violates (2018: 9).

On the basis of the above elements —experimental setup and assumptions— the F-R argument proceeds as follows. First, in order to compute the probability that the measurements of *X* and *Y* yield the results $\text{ok}_X$ and $\text{ok}_Y$, respectively, the state described by eq. (1) must be expressed as:

$$|\Psi\rangle = \frac{1}{\sqrt{12}}|\text{ok}_X\rangle|\text{ok}_Y\rangle - \frac{1}{\sqrt{12}}|\text{ok}_X\rangle|\text{fail}_Y\rangle + \frac{1}{\sqrt{12}}|\text{fail}_X\rangle|\text{ok}_Y\rangle + \sqrt{\frac{3}{4}}|\text{fail}_X\rangle|\text{fail}_Y\rangle \qquad (4)$$

From this eq. (4) it is clear that the probability of obtaining $\text{ok}_X$ and $\text{ok}_Y$ is 1/12.

The second part of the argument consists in showing that the observers involved in the experiment can draw a conclusion different from the above one on the basis of the following reasoning.[2] Let us consider the probability that $F_2$ obtains $-1/2$ in her $S_z$ measurement and $W_1$ obtains $\text{ok}_X$ in her *X* measurement; in order to compute this probability, the state described by eq. (1) must be expressed as:

$$|\Psi\rangle = \sqrt{\frac{2}{3}}|\text{fail}_X\rangle|\Downarrow\rangle + \frac{1}{\sqrt{6}}|\text{fail}_X\rangle|\Uparrow\rangle - \frac{1}{\sqrt{6}}|\text{ok}_X\rangle|\Uparrow\rangle \qquad (5)$$

From this eq. (5) it is easy to see that the considered probability is zero. Then, if $W_1$ obtains $\text{ok}_X$ in her *X* measurement on Lab $L_1$, she can infer that the outcome of $F_2$'s $S_z$ measurement on the qubit was $+1/2$. In turn, if $F_2$ obtains $+1/2$ in her $S_z$ measurement on the qubit, she can infer that the outcome of $F_1$'s *C* measurement on the quantum coin was *t*, because otherwise $F_1$ would send $F_2$ the qubit in state $|\downarrow\rangle$. And if $F_1$ obtains *t* in her *C* measurement on the quantum coin, she can infer that the outcome of $W_2$'s *Y* measurement on Lab $L_2$ will be $\text{fail}_Y$, because the outcome *t* is perfectly correlated with the state $|\Rightarrow\rangle$ of lab $L_2$, and $|\Rightarrow\rangle = |\text{fail}_Y\rangle$ (see eq. (3)). Therefore, from a nested reasoning it can be concluded that, when $W_1$ gets $\text{ok}_X$, she can infer that $W_2$ certainly gets $\text{fail}_Y$. But this conclusion contradicts what was inferred from eq. (4), that is, that there is a non-zero probability that $W_1$ gets $\text{ok}_X$ and $W_2$ gets $\text{ok}_Y$.

The reactions to the F-R argument have been multiple and varied. An interesting response emphasizes an implicit assumption of the argument: the non-relational view of quantum mechanics is an indispensable premise of the derivation. This is the view of Časlav Brukner, who considers, from an operational perspective, that the self-consistency condition (C) is too restrictive, since "the states referring to outcomes of different observers in a Wigner-friend type of experiment cannot be defined without referring to the specific experimental arrangements of the observers, in agreement with Bohr's idea of contextuality" (Brukner 2018: 8). From a non-operational standpoint, Dennis Dieks (2019) advocates, in the line of Carlo Rovelli's relational view (1996), for a perspectivalist interpretation of quantum mechanics, according to which more than one state can be assigned to the same physical system: the state and physical properties of a system are different in relation to different reference systems; when the perspectival nature of quantum states is included as a premise, no contradiction can be inferred from the F-R argument. According to Richard Healey (2018), the F-R argument implicitly depends on an inconclusive additional assumption, "intervention insensitivity", which guarantees that the truth-value of the outcome of a counterfactual measurement is insensitive to the occurrence of a physically isolated intervening event.

---

[2] We thank Jeffrey Bub for suggesting us this clear explanation of this part of the argument.

After supplying his clear and elegant reconstruction of the F-R argument as appeared in the 2016 paper, Jeffrey Bub (2018) claims that what he calls the "Frauchiger-Renner contradiction" shows that quantum mechanics should be understood probabilistically, as a new sort of non-Boolean probability theory, rather than representationally, as a theory about the elementary constituents of the physical world and how these elements evolve dynamically over time. In resonance with his information-theoretic interpretation of quantum mechanics, Bub conceives quantum mechanics formulated in Hilbert space as fundamentally a theory of probabilistic correlations that are structurally different from the correlations that arise in Boolean theories. Analogously to special relativity, as a theory about the structure of space-time that provides an explanation for length contraction and time dilation through the geometry of Minkowski space-time with no dynamical considerations, "[q]uantum mechanics, as a theory about randomness and nonlocality, provides an explanation for probabilistic constraints on events through the geometry of Hilbert space, but that's as far as it goes." (Bub 2018: 3).

From a completely different perspective, the conclusion of the F-R argument was rejected on the basis of Bohmian mechanics, the paradigmatic one-world no-collapse quantum theory. For instance, Anthony Sudbery (2017) offers a Bell-Bohmian reconstruction of the argument, claiming that it supplies a counter-example to the conclusion obtained by Frauchiger and Renner. With a similar reasoning, Dustin Lazarovici and Mario Hubert (2018) assert that any Bohm-type theory provides a logically consistent description of F-R *Gedankenexperiment* if the state of the entire system and the effects of all measurements are taken into account.

Since our discussion will be centered on the second part of the F-R argument, let us write it in a more concise form:

(a)  If $W_1$ gets $ok_X$, then she knows that $F_2$ got $+1/2$.

(b)  If $F_2$ gets $+1/2$, then she knows that $F_1$ got $t$.

(c)  If $F_1$ gets $t$, then she knows that $W_2$ will get $fail_Y$.

(d)  If $W_1$ gets $ok_X$, then she knows that $W_2$ gets $fail_Y$.

where (a), (b), and (c) are the premises of the reasoning, and (d) is its conclusion. In the following sections we will analyze this second part of the F-R argument from different perspectives.

## 3.- What does transitivity mean?

The first issue that arises when one faces the consistency condition is the question of what system of logic underlies the second part of the F-R argument. They seem to use a "folk" logic that makes plausible to infer the conclusion (d) from the three premises (a), (b), and (c). But, as it is already well known, this intuitive strategy is very dangerous in the quantum domain.

But the situation is even more confusing. In fact, the reasoning involves the application of a kind of transitivity according to which, from the proposition

If $W_1$ gets $ok_X$, then $W_1$ knows that $F_2$ knows that $F_1$ knows that $W_2$ gets $fail_Y$

it can be inferred that

If $W_1$ gets $ok_X$, then $W_1$ knows that $W_2$ gets $fail_Y$

This requires assuming that the following inference is valid:

'$A$ knows that $B$ knows that $C$ knows $p$' implies that '$A$ knows $p$' (6)

Surely somebody made Frauchiger and Renner to notice that they were not using a classical transitivity inference rule, because in a footnote of the published paper they stress: "Assumption (C) has some vague similarities with a transitivity relation. However, although the expression «knows that» indeed defines a binary relation, it is not transitive (for its domain and codomain are different sets)." (2018: 7). But if the argument does not rely on transitivity, which inference rule allows us to accept inference (6) as valid?

The question about the validity of inference (6) is relevant because, as stressed in logics, the verb 'to know' (as other verbs such as 'to believe', 'to hope', 'to hate', etc.) expresses a propositional attitude that generates an opaque context, that is, a linguistic context in which not always co-referential terms can be substituted *salva veritate*. For instance, although 'Lewis Carroll is Charles Dodgson', it may happen that the proposition 'John knows that Lewis Carroll was the author of *Alice in Wonderland*' is true, but the proposition 'John knows that Charles Dodgson was the author of *Alice in Wonderland*' is false. In turn, it may happen that 'John knows that Mary knows her passport number' is true, but 'John knows Mary's passport number' is false. In our case, from obtaining $t$ in her $C$ measurement, $F_1$ knows that the state of Lab $L_2$ will be $|\Rightarrow\rangle$; but even if $W_1$ knows that $F_1$ knows that the state of Lab $L_2$ will be $|\Rightarrow\rangle$, $W_1$ may ignore that the state of Lab $L_2$ will be $|\text{fail}_Y\rangle$, in spite of the fact that if $|\Rightarrow\rangle = |\text{fail}_Y\rangle$.

These logical considerations do not intend, per se, to dispute the validity of the F-R argument, but rather point to the need for reformulating it in a more precise form. In particular, the argument can be expressed with no reference to what the involved subjects know, but in terms of what the observers get in their measurements. For instance, Bruckner, although still talking about "«collapsing» others' knowledge into $W$'s knowledge" (2018: 8), and considering that the F-R argument "points to the necessity to differentiate between ones' knowledge about direct observations and ones' knowledge about others' knowledge that is compatible with physical theories" (2018: 8), reconstructs the argument under the following form (with the necessary terminology adjustment):

(a') If $W_1$ sees $\text{ok}_X$, then $F_2$ sees $+1/2$.

(b') If $F_2$ sees $+1/2$, then $F_1$ sees $t$.

(c') If $F_1$ sees $t$, then $W_2$ sees $\text{fail}_Y$.

(d') If $W_1$ sees $\text{ok}_X$, then $W_2$ sees $\text{fail}_Y$.

In this case, any reference to knowledge has vanished:[3] the three premises (a'), (b'), and (c') are conditional propositions, and the conclusion (d') is obtained by applying the classical inference rule of the transitivity of conditional.

With this reformulation of the second part of the F-R argument we are in a better position than in the previous case. However, it is not completely clear yet why we should accept the truth of the three premises. In particular, does the fact that the observers get precise values in measurements presuppose collapse?

### 4.- Single outcome versus collapse

The Frauchiger and Renner's article is confusing enough as to make difficult to decide at first sight whether the argument requires collapse or not. Several authors claim that the F-R argument does not include the hypothesis of collapse as one of its assumptions, and a significant part of its conceptual value relies on this fact. For example, Dieks (2019) understands the argument as based on unitary evolution for the dynamics of the quantum state, also during the measurement process. Bub, in turn, notes that the formalism used by Frauchiger and Renner does not presuppose "a suspension of unitary evolution in favor of an unexplained «collapse» of the quantum state." (Bub 2018: 2).

---

[3] Although on the basis of the 2016 version, Bub (2018) also offers a clear and concise reconstruction of the F-R argument that does not appeal to the knowledge of the agents involved in the experiment.

However, not everybody agrees with this view. For instance, Franck Laloë considers that the argument illustrates no inconsistency in quantum mechanics, but only the well-known fact that "the exact point at which the von Neumann reduction postulate should be applied is ill defined." (Laloë 2018: 1). Mateus Araújo (2018), in turn, finds "the flaw in Frauchiger and Renner's argument" in the fact that the predictions that Frauchiger and Renner claim to follow from quantum mechanics can only be obtained when collapse is added. These opinions are not completely unfounded on the basis of the article's content. In fact, the entire explanation of the "Wigner's friend paradox" included in the Introduction of the 2018 article presupposes collapse in measurements, and the new *Gedankenexperiment* is presented as an extension of Wigner's one, without pointing out any other difference. Moreover, all along the development of the no-go theorem, the hypothesis of collapse is not discussed, not even mentioned. After a long discussion with Renner, Araújo (2018: Update) concluded that Renner thinks that the assumption of collapse in measurement is just a part of quantum mechanics, so it doesn't need to be stated separately.

In the light of this divergence of opinions, the first point to emphasize is that the single-world assumption (S) does not amount to nor implies collapse. Whereas the hypothesis of collapse imposes the non-unitary modification of the system's state due to measurement, (S) says nothing about the system's state. (S) only establishes that any measurement has a single outcome, and this may happen even if the system persists in its unitary evolution, as in the case, for example, of the modal interpretations (see Lombardi and Dieks 2017). Once this is clearly understood, it can be formally proved that the F-R argument does not require the hypothesis of collapse to reach its conclusion.

In order to develop the proof, first let us clean the discourse of any reference to observers and what they know or see, since quantum mechanics, as a physical theory, does not talk about the mental or visual states of agents. For this purpose, we will reformulate the F-R argument in terms of quantum propositions of the form 'the property *P* has the value *p*', which will be represented as '*P* : *p*' for conciseness. From now on, we will use the symbols '¬', '∧', '∨', '→', and '↔' for negation, conjunction, disjunction, conditional, and biconditional respectively, as usual.

On the basis of these clarifications, the first part of the F-R argument leads to the conclusion that the following proposition can be asserted:

$$X:\text{ok}_X \wedge Y:\text{ok}_Y \tag{7}$$

On the other hand, the second part of the F-R argument is a reasoning that reads:

(a'') $X:\text{ok}_X \to S_z:+1/2$

(b'') $S_z:+1/2 \to C:t$

(c'') $C:t \to Y:\text{fail}_Y$

(d'') $X:\text{ok}_X \to Y:\text{fail}_Y$

By defining $\to$ in terms of $\wedge$, and by considering that $\text{fail}_Y$ and $\text{ok}_Y$ are the two only eigenvalues of $Y$, conclusion (d'') can be expressed as:

$$X:\text{ok}_X \to Y:\text{fail}_Y \equiv \neg(X:\text{ok}_X \wedge \neg Y:\text{fail}_Y) \equiv \neg(X:\text{ok}_X \wedge Y:\text{ok}_Y) \tag{8}$$

The contradiction of the F-R argument results from eqs. (7) and (8). QED

Now, the F-R argument is formulated in a sufficiently clear way so as to formally prove that the involved propositions can be asserted without assuming collapse, but only accepting assumption (Q): a quantum proposition can be asserted/denied when the Born rule assigns probability 1/0 to it. The proof requires recalling that the eigenstates/eigenvalues of the observable $A$ of lab $L_1$ are correlated with the eigenstates/eigenvalues of the observable $C$ of the coin, and the eigenstates/eigenvalues of the observable $B$ of lab $L_2$ are correlated with the eigenstates/eigenvalues of the observable $S_z$ of the qubit:

$$C:h \leftrightarrow A:H \qquad C:t \leftrightarrow A:T \tag{9}$$

$$S_z:+1/2 \leftrightarrow B:\Uparrow \qquad S_z:-1/2 \leftrightarrow B:\Downarrow \tag{10}$$

- Proposition (7) can be asserted in some situations because $\Pr(X:\text{ok}_X \wedge Y:\text{ok}_Y)=1/12$, which is implied by eq. (4).

- Proposition (a'') can be transformed by taking into account eqs. (10), the definition of $\to$ in terms of $\wedge$, and the fact that $\Uparrow$ and $\Downarrow$ are the only two eigenvalues of $B$:

$$X:\text{ok}_X \to S_z:+1/2 \equiv X:\text{ok}_X \to B:\Uparrow \equiv \neg(X:\text{ok}_X \wedge \neg B:\Uparrow) \equiv \neg(X:\text{ok}_X \wedge B:\Downarrow) \tag{11}$$

In order to assert $\neg(X:\text{ok}_X \wedge B:\Downarrow)$, it must be proved that $\Pr(X:\text{ok}_X \wedge B:\Downarrow)=0$, which in turn requires to express the state $|\Psi\rangle$ in the basis $X$-$B$ of $\mathcal{H}_1 \otimes \mathcal{H}_2$ as follows:

$$|\Psi\rangle = \sqrt{\frac{2}{3}}|\text{fail}_X\rangle|\Downarrow\rangle + \frac{1}{\sqrt{6}}|\text{fail}_X\rangle|\Uparrow\rangle - \frac{1}{\sqrt{6}}|\text{ok}_X\rangle|\Uparrow\rangle \tag{12}$$

- Proposition (b'') can be transformed by taking into account eqs. (9) and (10), again the definition of $\to$ in terms of $\wedge$, and the fact that $T$ and $H$ are the two only eigenvalues of $A$:

$$S_z : +1/2 \rightarrow C:t \equiv B:\Uparrow \rightarrow A:T \equiv \neg(B:\Uparrow \wedge \neg A:T) \equiv \neg(B:\Uparrow \wedge A:H) \qquad (13)$$

In order to assert $\neg(B:\Uparrow \wedge A:H)$, it must be proved that $\Pr(B:\Uparrow \wedge A:H) = 0$, which in turn requires to express the state $|\Psi\rangle$ in the basis *B-A* of $\mathcal{H}_1 \otimes \mathcal{H}_2$ as follows:

$$|\Psi\rangle = \frac{1}{\sqrt{3}}|H\rangle|\Downarrow\rangle + \frac{1}{\sqrt{3}}|T\rangle|\Uparrow\rangle + \frac{1}{\sqrt{3}}|T\rangle|\Downarrow\rangle \qquad (14)$$

- Proposition (c'') can be transformed by taking into account eqs. (9), again the definition of $\rightarrow$ in terms of $\wedge$, and the fact that $\text{fail}_Y$ and $\text{ok}_Y$ are the two only eigenvalues of *Y*:

$$C:t \rightarrow Y:\text{fail}_Y \equiv A:T \rightarrow Y:\text{fail}_Y \equiv \neg(A:T \wedge \neg Y:\text{fail}_Y) \equiv \neg(A:T \wedge Y:\text{ok}_Y) \qquad (15)$$

In order to assert $\neg(A:T \wedge Y:\text{ok}_Y)$, it must be proved that $\Pr(A:T \wedge Y:\text{ok}_Y) = 0$, which, in turn, requires to express the state $|\Psi\rangle$ in the basis *A-Y* of $\mathcal{H}_1 \otimes \mathcal{H}_2$ as follows:

$$|\Psi\rangle = \frac{1}{\sqrt{6}}|H\rangle|\text{fail}_Y\rangle + \frac{1}{\sqrt{6}}|H\rangle|\text{ok}_Y\rangle + \sqrt{\frac{2}{3}}|T\rangle|\text{fail}_Y\rangle \qquad (16)$$

Summing up, the propositions involved in the F-R argument can be inferred from the formalism of standard quantum mechanics without appealing to the hypothesis of collapse or to any other assumption about measurement. The only "trick" is to bring into play cases of probability equal to zero or to one.

Let us recall that in the original Wigner's friend argument, the paradox arises when comparing the collapsed state of the friend inside the lab and the superposition assigned by Wigner from the outside. If the conclusion of the F-R argument depended on collapse, it would lose much of its appealing since it would offer no much novelty when compared with the original Wigner's friend argument and would depend on an interpretive assumption. By contrast, what has shocked most of the physics community is that the argument seems to show an internal inconsistency of quantum mechanics at the level of probabilities, independently of any interpretive addition.

## 5.- Using classical logic in a quantum context

Up to this point we have seen that the F-R argument leads to a contradiction without appealing to observers' states of consciousness, memory, or knowledge, and without introducing the collapse hypothesis. This suggests that we are facing a really new and powerful no-go theorem. However, there are good reasons to suspect that, although the theorem is powerful, it is not substantially new.

In fact, in the discussions around the argument, few authors stress with sufficient strength that it is based on inferences belonging to classical logic. An exception is Bruckner, who, still talking about knowledge instead of about quantum propositions and without a full proof, points out that "«collapsing» others' knowledge into *W*'s knowledge […] is equivalent in its implications to considering all the statements as belonging to a single Boolean algebra" (2018: 8). The detailed reconstruction of the previous section makes easy to see why. In the reasoning of the second part of the F-R argument, the conclusion (d'') is obtained by applying the classical inference rule of transitivity of conditional, according to which, from $(p \rightarrow q) \wedge (q \rightarrow r)$, $(p \rightarrow r)$ can be inferred. Therefore, it is necessary to make the conjunction between the conditional propositions (a''), (b''), and (c'') in order to obtain (d'') by transitivity. Each one of those propositions was obtained from a probabilistic assertion that was computed by expressing the state $|\Psi\rangle$ in a different basis of the Hilbert space $\mathcal{H}_1 \otimes \mathcal{H}_2$ of the system composite of the two labs $L_1$ and $L_2$: the bases *X-B*, *B-A*, and *A-Y* for (a''), (b''), and (c'') respectively (see eqs. (12), (14), and (16) respectively). But they are three different bases, rotated with respect to each other.[4] In other words, arriving at the contradiction by means of the F-R argument requires making classical conjunctions between propositions corresponding to different contexts, something that the non-Boolean structure of the quantum propositions forbids, as it is well-known since 1967, when Simon Kochen and Ernst Specker demonstrated their famous theorem.

Let us recall that the F-R argument is built on three assumptions: it would show that accepting (Q), (C), and (S) leads to a contradiction. In which of the three assumptions is the admissibility of conjunctions between propositions corresponding to different contexts included? Given the content of (Q) and (S), it seems plausible that such a logical admissibility is included in assumption (C), which "demands consistency, in the sense that the different agents' predictions are not contradictory" (Frauchiger and Renner 2018: 2). So, let us focus our attention on it.

According to Frauchiger and Renner (2018: 7, caption of Figure 3), "If a theory T (such as quantum theory) enables consistent reasoning (C) then it must allow any agent A to promote the conclusions drawn by another agent A' to his own conclusions, provided that A' has the same initial knowledge about the experiment and reasons within the same theory T." This means that, if the shared initial knowledge is accepted, (C) requires that two agents (i) reason with the same theory T, and (ii) use the same system of logic to obtain their conclusions. But, what is the relation between a theory T —in this case, a physical theory as quantum mechanics— and the system of logic by means of which the agents draw their own conclusions? One alternative is to consider that the theory T is constituted by a mathematical structure and some postulates, and the logic by means of which the agents make inferences on the basis of the theory is classical, defined on a Boolean structure of propositions. This is the strategy followed by Frauchiger and Renner, who rely on the mathematical structure of quantum mechanics and the Born Rule as one of its postulates, but allow the agents to use classical logic to make their inferences. On this basis, they prove that accepting the assumptions (Q) —compliance with T=quantum theory—, (S) —single outcomes of measurements or single truth values of propositions—, and (C) —that the agents make inferences on the basis of T=quantum theory with classical logic— leads to a contradiction. But this is a direct consequence of the Kochen-Specker theorem. Therefore, according to this view, the F-R argument is an original and interesting way to get a result already obtained by other means. In other words, the argument is a no-go theorem, but what does not go was already well known and supplies no new knowledge about quantum mechanics —and even less offers a proof of an internal inconsistence of the theory.

---

[4] The fact that the bases *X-B*, *B-A*, and *A-Y* are different can be proved by defining three observables $O_{XB}$, $O_{BA}$, and $O_{AY}$ acting on $\mathcal{H}_1 \otimes \mathcal{H}_1$, whose eigenvectors are the members of the bases *X-B*, *B-A*, and *A-Y* respectively, and by proving that those three observables do not commute with each other.

However, there is another alternative regarding how to conceive the relation between a theory T and the system of logic by means of which inferences are made. In fact, it can be considered that a physical theory T constrains the range of systems of logic that can be used to make inferences with its propositions. In other words, the mathematical structure of the theory T embodies the algebra of T-propositions, which restricts the admissible logic to operate with those propositions. Therefore, in the case of quantum mechanics, making inferences with classical logic on quantum propositions is not legitimate, because it is in conflict with the non-Boolean algebra of those propositions. This would be the position taken by a professor who, in an exam on quantum mechanics, rejected a student's answer that includes a conjunction of propositions corresponding to the values of non-commuting observables. From this viewpoint, the F-R argument is illegitimate, to the extent that it assumes compliance with quantum theory by (Q), but according to (C) allows the agents to make inferences with classical logic.

There are, then, two alternatives to assess the F-R argument: legitimate but not new, or perhaps new but not legitimate. Although we have our own preference for one of the two alternatives, we let the readers free to make their decisions. But what seems quite clear is that, in neither of the two cases the F-R argument provides a result that shakes the foundations of quantum mechanics.

## 6.- Conclusions

The wide and strong impact of the F-R argument is undeniable, not only for the high number of comments that appeared under different forms since its first presentation in 2016, but also for the severe consequences for the foundations of the theory that it supposedly involves. In the foundations of physics community, the argument has been largely discussed from different viewpoints and on the basis of very different interpretations of the proposal. The disagreements about which the assumptions actually are and what the argument really proves are a manifestation of the fact that the presentation of the argument might be clearer than it is.

In this brief article we have analyzed the F-R argument, with the purpose of offering a detailed reconstruction that can be helpful for future discussions. On the basis of this reconstruction we have shown that the argument can be formulated only in terms of quantum propositions, in such a way that any ambiguity or confusion derived from introducing what subjects know or see in the reasoning can be avoided. In addition, our reconstruction has allowed us to prove, in a precise formal way, that the argument does not require the hypothesis of collapse to arrive to its conclusion: the propositions that take part of the argument are asserted/denied when the Born rule assigns probability 1/0 to them, without supposing that the state evolves non-unitarily in the measurement process. Finally, we have shown that the contradiction resulting from the F-R argument is inferred by making classical conjunctions between different and incompatible contexts, a strategy that stands in conflict with the well-known contextuality of quantum mechanics derived from the non-Boolean structure of quantum propositions. This fact leaves us with two alternatives, depending on how we conceive the relation between a physical theory T and the system of logic supporting inferences on T-propositions: either the F-R argument is an original way to reproduce the proof of the contextuality of quantum mechanics, or the argument is illegitimate because appeals to inferences forbidden by the algebraic structure of quantum propositions. In both cases, the F-R argument lacks the high conceptual relevance suggested by its great impact under the form of comments, discussions, and even alarmist claims about the "breaking" or the "inconsistence" of quantum mechanics.

**Acknowledgements**: We are extremely grateful to Dennis Dieks for encouraging us to write down these ideas, and to Jeffrey But for his enlightening comments to a previous version of the present article.

events/eth-news/news/2018/09/errors-in-the-quantum-world.html. See also *Science Daily*, September 18, 2018, avaliable at www.sciencedaily.com/releases/2018/09/180918114438.htm.